\documentclass[letterpaper,10pt]{article}

\usepackage{opex3}
\usepackage{hyperref}
\usepackage{graphicx}
\usepackage{dcolumn}
\usepackage{bm}
\usepackage{amsmath}
\usepackage{amsfonts}
\usepackage{amssymb}
\usepackage{appendix}
\newcommand{\gper}{\gamma_\perp}
\newcommand{\gpar}{\gamma_\parallel}

\begin{document}

\title{Steady-State Ab Initio Laser Theory for N-level Lasers}
\author{Alexander Cerjan,$^1$ Yidong Chong,$^1$ Li Ge,$^2$ and A. Douglas Stone$^{1*}$}
\address{${}^1$Department of Applied Physics, Yale University, New Haven, CT
06520}
\address{${}^2$Department of Electrical Engineering, Princeton University, Princeton, NJ
08544}
\address{${}^*$Corresponding author: douglas.stone@yale.edu}
\date{\today}

\begin{abstract}
We show that Steady-state Ab initio Laser Theory (SALT) can be applied
to find the stationary multimode lasing properties of an $N$-level
laser.  This is achieved by mapping the $N$-level rate equations to an
effective two-level model of the type solved by the SALT algorithm.
This mapping yields excellent agreement with more computationally
demanding $N$-level time domain solutions for the steady state.
\end{abstract}


%\maketitle

\section{Introduction}

Semiclassical laser theory, which neglects the quantum fluctuations of
the electromagnetic field, is widely used to describe and simulate
lasers \cite{haken,lamb63}.  In principle, it correctly describes the
laser thresholds and frequencies, the spatial pattern of the lasing
modes, and the laser output power, including all classical non-linear
effects, such as spatial hole-burning, gain saturation, and mode and
phase locking.  Essentially the theory describes Maxwell's equations
in an open cavity, coupled to the non-linear polarization of the gain
medium.  The gain polarization can be described using either a
classical non-linear oscillator model \cite{siegman}, or a
quantum-mechanical model of $N$ atomic levels in which the
polarization and level populations obey the equations of motion of the
quantum density matrix.  The simplest version of the theory, used
widely in textbooks, is the two-level Maxwell-Bloch (MB) model
\cite{haken}; however, most design and characterization simulations of
lasers use models with $N = 3$ or more levels.  In addition, most
theoretical solutions for the semiclassical laser equations employ a
large number of simplifying assumptions in order to make them
analytically tractable, most notably neglecting the openness of the
cavity and/or treating only simple one-dimensional (1D) or ring
cavities, as well as approximating the non-linear interactions to
cubic order. The results are typically not useful for quantitative
modeling.  Until recently, the only useful way to obtain quantitative
results for non-trivial laser structures was to integrate the
semiclassical laser equations in space and time.  For novel and
interesting new laser structures with non-trivial 2D and 3D cavity
geometries, such simulations are at the limits of computational
feasibility, making it difficult to study a large parameter space or
ensemble of designs.

In the past five years a new approach to finding the stationary
solutions of the semiclassical laser equations has been developed,
known as Steady state Ab initio Laser Theory
(SALT)\cite{salt1,salt2,saltsci,spasalt}.  SALT treats the openness of
the cavity exactly, and the multimode non-linear interactions to
infinite order (with two approximations, to be discussed below).  It
is applicable to cavities of arbitrary complexity in 2D and 3D,
although we will discuss here only the scalar wave equation coupled to
a gain medium.  Importantly, this approach eliminates the need to
perform a time integration to steady-state, dramatically reducing the
computational effort and allowing one to study cavities with high
spatial complexity, such as 2D random lasers \cite{saltsci,cao},
photonic crystal lasers \cite{phot,pcsel} and chaotic disk microlasers
\cite{disk,li_thesis}.

SALT was originally formulated to solve the steady-state two-level
Maxwell-Bloch equations \cite{salt1,salt2} in the standard slowly-varying
envelope approximation (SVEA), and an iterative algorithm was
developed \cite{salt2} to solve the resulting SALT equations.
Subsequently it was realized that the SVEA afforded no advantage in
numerical solutions of the SALT equations, so this approximation was
dropped, leading to slightly different SALT equations
\cite{li_thesis,tandy}, which were used in all subsequent work
\cite{saltsci,spasalt,pcsel}.  Recently an important generalization of
the SALT solution algorithm was developed, improving its performance
for laser cavities with complex spatial index variation and
inhomogeneous pumping \cite{spasalt}.  In its present form, SALT only
employs two approximations, the stationary inversion approximation
(SIA) and the rotating wave approximation (RWA), which are both well
satisfied for most lasers of interest.  In Ref.~\cite{tandy}, the
results of SALT were compared to the full time-dependent solution of
the MB equations for a simple 1D cavity in the multimode regime, well
above threshold.  Excellent agreement was found in the parameter
regime for which the SIA holds.  This was, to our knowledge, the first
demonstration of a frequency-domain method which agrees with exact
time-domain methods for above-threshold multimode lasing.

Previous applications of SALT have focused on two-level gain media.
In order for SALT to be a useful modeling tool, it is necessary to
demonstrate that it can be applied to $N$-level lasers.  In the
current work, we show analytically that the steady-state equations for
an $N$-level laser can be reduced to those for an effective two-level
system, and hence solved using the efficient SALT algorithm with
essentially the same degree of computational effort.  We also explore
how this effective two-level system differs from the ordinary
two-level laser.  Next, we present a numerical comparison between the
results of SALT calculations and exact $N$-level finite-difference
time-domain (FDTD) calculations, for the same simple 1D laser studied
in \cite{tandy}, as well as for a 1D random laser.  We note that a
similar comparison between SALT and FDTD has been performed for a
four-level high-Q single-mode photonic crystal laser in Ref.~\cite{pcsel}, with
good agreement found.  Here, we test SALT's accuracy in treating the
more challenging case of multimode lasing in low-Q and random
cavities.

\section{Effective two-level systems}

\subsection{Four level system analysis}

\begin{figure}[!h]
\centering
\includegraphics[width=0.7\textwidth]{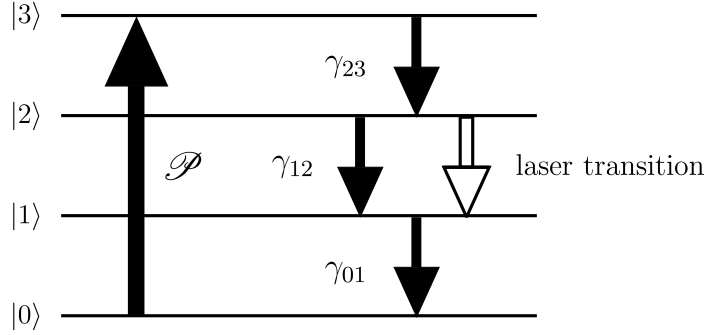}
\caption{Schematic of a four-level gain medium. \label{schem}}
\end{figure}

We illustrate the approach using the semi-classical laser equations
\cite{haken} for the four-level atomic gain medium shown in
Fig.~\ref{schem}:
\begin{eqnarray}
4\pi \ddot{\mathbf{P}}^+ &=& c^2 \nabla^2 \mathbf{E}^+ - \varepsilon_c(\mathbf{r})
\ddot{\mathbf{E}}^+ \label{waveqn} \\
\dot{\mathbf{P}}^{+} &=& -\left(i\omega_a + \gamma_{\perp}\right)\mathbf{P}^{+} + 
\frac{g^2}{i \hbar}\mathbf{E}^+\left(\rho_{22} - \rho_{11}\right) \label{poleqn} \\
\dot \rho_{33} &=& \mathcal{P} \left( \rho_{00} - \rho_{33} \right) - \gamma_{23}
\rho_{33} \label{rbeqn} \\
\dot \rho_{22} &=& \gamma_{23} \rho_{33} - \gamma_{12} \rho_{22} - 
\frac{1}{i\hbar}\mathbf{E}^+ \cdot \left((\mathbf{P}^{+})^* - \mathbf{P}^{+} \right) \\
\dot \rho_{11} &=& \gamma_{12} \rho_{22} - \gamma_{01} \rho_{11} +
\frac{1}{i\hbar}\mathbf{E}^+ \cdot \left((\mathbf{P}^{+})^* - \mathbf{P}^{+} \right) \\
\dot \rho_{00} &=& \gamma_{01} \rho_{11} - \mathcal{P} 
\left( \rho_{00} - \rho_{33} \right). \label{reeqn}
\end{eqnarray}
The RWA has already been made, and we have assumed a lasing structure
with one or two directions of translational symmetry, so that TM and
TE polarizations are conserved and Maxwell's equations reduce to a
scalar Helmholtz equation \cite{TE}.  $\mathbf{E}^+$ and
$\mathbf{P}^+$ are the positive frequency components of the scalar
electric and polarization fields respectively.  $\rho_{ii}$ is the
population density of level $|i \rangle$, $\omega_a$ is the frequency
of the gain center, $\gamma_\perp$ is the gain width (polarization
dephasing rate), $g$ is the dipole matrix element, $\mathcal{P}$ is
the pump rate, and $\gamma_{ij}$ is the decay rate from level $|j
\rangle$ to level $|i \rangle$.  The four levels are labelled from $0
- 3$ in order of increasing energy (Fig.~\ref{schem}).  The
polarization equation (\ref{poleqn}) is obtained from the four-level
density matrix equation of motion, assuming that only the level $2 \to
1$ transition will be inverted and lase.  Often in FDTD calculations a
real classical oscillating dipole equation is used to describe the
polarization \cite{pcsel}; in Appendix A, we show that this yields
essentially the same results, with an appropriate identification of
parameters.  In the rate equations (\ref{rbeqn})-(\ref{reeqn}), the
pump is coherently acting between levels $|0\rangle$ and $|3\rangle$.

The polarization equation (\ref{poleqn}) incorporates the inversion
$D(x,t) \equiv \rho_{22} (x,t) - \rho_{11} (x,t)$, similar to the
polarization equation for a two-level laser.  By assuming that the
non-lasing populations are stationary, $\dot\rho_{00} = \dot\rho_{33}
= 0$, we can show that $D(x,t)$ obeys
\begin{equation}
\dot D = -\gamma_{\parallel}'(D - D_0') - \frac{2}{i\hbar}\mathbf{E}^+
\cdot \left((\mathbf{P}^{+})^* - \mathbf{P}^{+} \right), \label{efinv}
\end{equation}
which is precisely the form of the inversion equation for the
two-level medium \cite{haken,fu}.  The parameters $D_0'$ and
$\gamma_\parallel'$ serve as an effective equilibrium inversion and inversion
relaxation rate respectively, and are given by \cite{li_thesis}:
\begin{eqnarray}
\gamma_\parallel' &=& 2 \gamma_{12} \left( 1 + \frac{S}{2 +
  \frac{\gamma_{01}}{ \mathcal{P}} + 2
  \frac{\gamma_{01}}{\gamma_{23}}}\right) \label{eq8} \\ D_0' &=&
\frac{S \mathcal{P} \, n}{\gamma_{01} + \left(S + 2 +
  2\frac{\gamma_{01}} {\gamma_{23}} \right) \mathcal{P}}, \label{eq9}
\end{eqnarray}
where $S = (\gamma_{01} - \gamma_{12})/\gamma_{12}$ and $n = \sum_i
\rho_{ii}$ is the total density of gain atoms.  Eq.~(\ref{eq9}) for the
inversion in the absence of laser emission (which here acts as the
effective pump parameter) has been discussed by Siegman
\cite{siegman}, while Eq.~(\ref{eq8}) for the effective relaxation
rate has been derived for a special case by Khanin \cite{khanin}.
These expressions have not been used previously to solve the
four-level lasing equations in terms of the two-level solutions, as we
do here.  If we use an incoherent pump, the $ \mathcal{P} (\rho_{00} -
\rho_{33})$ terms in (\ref{rbeqn}) and (\ref{reeqn}) would be
replaced by $\mathcal{P}\rho_{00}$; the effect of this would be the
removal of the factor of $2$ preceding the ratio
$\gamma_{01}/\gamma_{23}$ in the denominators of (\ref{eq8}) and
(\ref{eq9}). For typical laser systems, each denominator is dominated
by the $\gamma_{01}/\mathcal{P}$ term, so the difference between
coherently and incoherently pumping the system is negligible
\cite{caveat}.

Given the parameters $\{\gamma_{01},\gamma_{12},\gamma_{23},
\mathcal{P}\}$ describing the four-level medium of Fig.~\ref{fig1}, we
can calculate the effective pump and relaxation rates, and feed those
(together with the cavity dielectric function, lasing transition
frequency $\omega_a$, and gain linewidth $\gamma_{\perp}$) into the
SALT algorithm.  That will yield the steady-state lasing properties of
this four-level laser.

\subsection{Arbitrary number of levels}

A general $N$-level laser can be treated via an analogous procedure.
Suppose we have a gain medium with an arbitrary number of levels, $N$.
We assume that there is only a single lasing transition, between two
levels which we denote by $|u\rangle$ and $|l\rangle$.  The population
of each of the $N-2$ non-lasing levels obeys
\begin{equation}
\dot \rho_{ii} = \sum_{j} \gamma_{ij}\rho_{jj} - \sum_{j} \gamma_{ji}
\rho_{ii} + \gamma_{iu} \rho_{uu} + \gamma_{il} \rho_{ll} -
(\gamma_{ui}+\gamma_{li})\rho_{ii},
\end{equation}
where the sums are taken over all of the non-lasing states and
$\gamma_{ij}$ is the rate at which atoms transition from state $|j
\rangle$ to $|i \rangle$ which is either a decay or pump rate
depending on the relative energies of the states.  In this way, we can
incorporate decay and pump processes between any levels.  We again
assume that $\dot{\rho}_{ii} = 0$ for all non-lasing levels.  Then
\begin{equation}
  \sum_j \left[(s_i + \gamma_{ui} + \gamma_{li}) \delta_{ij} -
    \gamma_{ij}\right] \rho_{jj} = \gamma_{iu} \rho_{uu} +
  \gamma_{il}\rho_{ll},
  \label{rhoii matrix}
\end{equation}
where $s_i = \Sigma_j \gamma_{ji}$ and $\delta_{ij}$ is the Kronecker
delta.  The term in brackets on the left hand side corresponds to an
$(N-2)\times(N-2)$ matrix, which we denote as $R$.  Upon inverting
Eq.~(\ref{rhoii matrix}) and substituting it into the equations of
motion for the lasing levels, we obtain
\begin{equation}
\gamma_\parallel' = \frac{B_l T_u - B_u T_l}{T_u + T_l}, \qquad\; D_0'
= \frac{B_u + B_l}{T_u + T_l} \; \frac{n}{\gamma_\parallel'},
\end{equation}
where
\begin{eqnarray}
  T_{u/l} &=& 1 + \sum_{ij} [R^{-1}]_{ij} \, \gamma_{j,u/l} \\
  B_u &=& - s_u + \sum_{ij} (\gamma_{ui} - \gamma_{li}) [R^{-1}]_{ij}
  \gamma_{ju}, \\ B_l &=& s_l + \sum_{ij} (\gamma_{ui} -
  \gamma_{li}) [R^{-1}]_{ij} \gamma_{jl}.
\end{eqnarray}
The details of this calculation are given in Appendix B.

\section{Physical Limits of Interest}

Returning to the typical four-level case, we take note of two
important physical regimes.  The first, the linear regime, is 
$\gamma_{23} \sim
\gamma_{01} \gg \gamma_{12} \gg \mathcal{P}$, for which one recovers
the expected behavior that the equilibrium inversion increases
linearly with the pump and that $\gamma_\parallel'$ is a constant:
\begin{eqnarray}
  \gamma_\parallel' &\approx& 2 \gamma_{12}, \label{l1} \\ D_0'
  &\approx& \frac{\mathcal{P}}{\gamma_{12}}n \label{l2}.
\end{eqnarray}
In this case, varying the equilibrium inversion and the pump strength
are essentially equivalent.

The second regime of interest, the non-linear regime, 
is $\gamma_{23} \sim \gamma_{01} \gg
\gamma_{12} \sim \mathcal{P}$, i.e.\ when the slow decay rate between
the lasing levels is on the same order as the pump rate.  In this
regime, $\gamma_\parallel'$ increases with increasing pump and $D_0'$
saturates with increasing pump:
\begin{eqnarray}
\gamma_\parallel' &\approx& 2 \left(\gamma_{12} + \mathcal{P}
\right), \label{gp1} \\ D_0' &\approx& \frac{1}{1+
  \frac{\mathcal{P}}{\gamma_{12}}} \left( \frac{\mathcal{P}}
      {\gamma_{12}} \right) n. \label{gp2}
\end{eqnarray}
This regime is also interesting from the viewpoint of SALT.  As
$\gamma_\parallel'$ increases with $\mathcal{P}$, a laser
could satisfy the inequality $\gamma_\parallel' \ll
\gamma_\perp$ near threshold, leading to stationary inversion and an
accurate solution via SALT, but fail to satisfy the inequality as the
pump becomes stronger, leading to a decrease in the accuracy of SALT.

For a system with an arbitrary number of levels, the first regime
always occurs at sufficiently small pump values; the second regime is
obtainable if electrons in the upper lasing level are relatively
long-lived compared to electrons in other levels.

\section{Brief Summary of SALT}

For completeness we briefly outline SALT. The $E^+$ and $P^+$ fields
are assumed to obey a multi-mode ansatz
\begin{align}
  \begin{aligned}
    E^+(\vec{r},t) &= \sum_{\mu=1}^M \Psi_\mu(\vec{r})\,e^{-i\omega_\mu t}, \\
    P^+(\vec{r},t) &= \sum_{\mu=1}^M p_\mu(\vec{r})\,e^{-i\omega_\mu t},
    \label{mode ansatz}
  \end{aligned}
\end{align}
where the indices $\mu = 1, 2, \cdots, M$ label the different lasing
modes, and the field and polarization are now explicitly scalar
quantities.  The total number of modes, $M$, is not given, but
increases in unit steps from zero as we increase the pump strength
$D_0$.  The values of $D_0$ at which each step occurs are the
(interacting) modal thresholds, to be determined self-consistently
from the theory. The real numbers $\omega_\mu$ are the lasing frequencies
of the modes (henceforth $c=1$), which will also be determined
self-consistently.

We insert the ansatz (\ref{mode ansatz}) into the two-level laser
equations, and employ the stationary inversion approximation (SIA)
$\dot{D} = 0$.  The result is a set of coupled nonlinear differential
equations, which are the fundamental equations of SALT \cite{spasalt}:
\begin{eqnarray}
  \left[\nabla^2 + \left(\epsilon_c(\vec{r}) + \frac{\gper
      D(\vec{r})}{k_\mu - k_a + i\gamma_\perp} \right)k_\mu^2\right]
  \Psi_\mu(\vec{r}) = 0, \label{TSG1} \\ D(\vec{r}) = D_0(\vec{r}) \,
  \left[1 +\sum_{\nu=1}^M \Gamma_\nu
    |\Psi_\nu(\vec{r})|^2\right]^{-1}. \label{TSG2}
\end{eqnarray}
$\Psi$ and $D$ are now dimensionless, measured in their natural units
$E_c = \hbar\sqrt{\gpar\gper}/(2g)$ and $D_c = \hbar\gper/(4\pi g^2)$,
and $\Gamma_\nu \equiv \gper^2/(\gper^2 + (\omega_\nu-\omega_a)^2)$ is the
Lorentzian gain curve evaluated at frequency $\omega_\nu$. Note that these
equations are time-independent; Eq.~(\ref{TSG1}) is a stationary wave
equation for the electric field mode $\Psi_\mu$, with an effective
dielectric function consisting of both the ``passive'' contribution
$\epsilon_c(\vec{r})$ and an ``active'' contribution from the gain
medium.  The latter is frequency-dependent, and has both a real part
and a negative (amplifying) imaginary part.  It also includes
infinite-order nonlinear ``hole-burning'' modal interactions, seen in
the $|\Psi_\nu|^2$ dependence of (\ref{TSG2}).  In addition, we make
the key requirement that $\Psi_\mu$ must be purely out-going outside
the cavity; it is this condition that makes the problem non-Hermitian.
It is worth noting that the SIA is not needed until at least two modes
are above threshold, so (\ref{TSG2}) is exact for single-mode lasing
up to and including the second threshold (aside from the well-obeyed
RWA).

These equations are solved efficiently by projecting them onto a
complete biorthogonal set of purely outgoing states with external
wavevectors, $k_\mu$, equal to the lasing frequencies.  We refer to
these states as the {\it threshold constant flux} (TCF) states,
because one member of the basis set is always equal to the
(non-interacting) threshold lasing mode, leading to very rapid
convergence of the basis expansion above threshold \cite{spasalt}.
The major computational effort in solving the SALT equations by this
approach is in calculating the (linear) TCF states.  The SALT
solutions are obtained for successive values of the pump increasing
from the first threshold; at each step, the coefficients of the lasing
modes in the TCF basis are obtained using a standard non-linear
solver, with the coefficients from the previous step as an initial
guess (which is never far from the correct solution).  Unlike FDTD, in
the current version of SALT one cannot simply directly solve at a
fixed pump value, well above threshold.  However, even with this
limitation, SALT is much more efficient, and provides substantial
physical insight, as we will discuss below.

\section{Numerical comparison}

\begin{figure}[!h]
\centering
\includegraphics[width=0.7\textwidth]{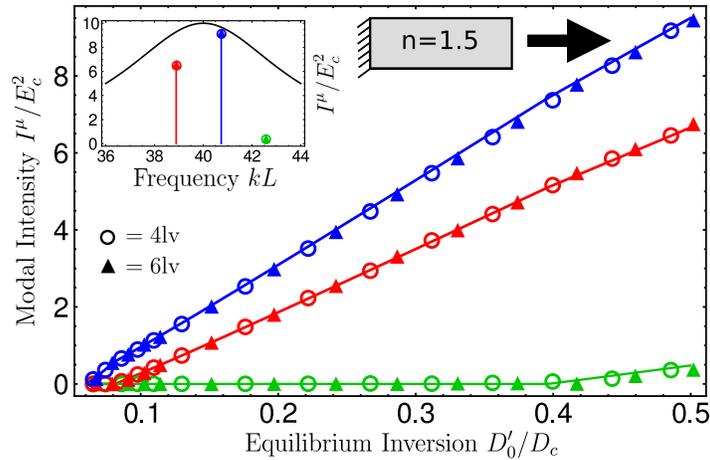}
\caption{Modal intensities as functions of the normalized equilibrium
  inversion $D_0' / D_{c}$ (effective pump) in a 1D microcavity edge emitting laser
  (schematic inset).  The cavity is bounded on one side by a perfect
  mirror and on the other side by air, and has uniform refractive
  index $n=1.5$.  Solid lines are results obtained by the
  time-independent SALT method; open circles are results of FDTD
  simulations with a coherently pumped four-level medium
  (Fig.~\ref{schem}); solid triangles are results of FDTD simulations
  with a coherently pumped six-level medium with a lasing transition
  between $|3 \rangle$ and $|1 \rangle$.  Simulation parameters are
  given in Appendix D.  Both the four-level and six-level media are
  chosen to satisfy SIA.  The dephasing rate is $\gamma_\perp = 4.0$.  The
  four-level system is in the linear regime described by
  Eq.~(\ref{l1})-(\ref{l2}).  The six-level system is calculated using
  the formula in the appendix B, but is in the non-linear regime
  described by Eq.~(\ref{gp1})-(\ref{gp2}).  The spectra at $D_0 / D_c
  = 0.488$, and the gain curve, are shown in the upper left inset. \label{fig1}}
\end{figure}

To perform a well-controlled comparison between SALT and the
four-level laser equations (\ref{waveqn})-(\ref{reeqn}), as well as
$N$-level generalizations, we studied 1D microcavity lasers for which
the FDTD calculations are tractable and fast enough to generate
extensive steady-state data.  We first consider the same simple edge
emitting uniform-index laser treated in
Refs.~\cite{salt1,salt2,tandy}, with a perfect mirror at the origin,
active region of length $L$ terminating abruptly in air (see
schematic, Fig.~\ref{fig1}). The simulations were carried out using
standard FDTD for the electromagnetic field, and Crank-Nicholson
discretization for the polarization and rate equations based on the
method of Bid\'{e}garay \cite{bid} (in which the polarization and
inversion are spatially aligned with the electric field but updated at
the same time steps as the magnetic field). The reported modal
intensities are calculated by Fourier transforming the electric field
at the cavity boundary after the simulation has reached steady state
(see \S\ref{efficiency section} for a discussion of the steady-state
criterion).  The lasing transition frequency $\omega_a$ is chosen so that
$n_0 k_a L = 60$, corresponding to roughly ten wavelengths of
radiation within the cavity.  Physical quantities are reported in
terms of their natural scales, $D_c = \hbar \gamma_\perp / (4 \pi
g^2)$ and $E_c = (\hbar/2g)\sqrt{\gamma_\perp
\gamma_\parallel'}$.  In addition, we take $c = \hbar =
1$ and measure rates in dimensionless units,
i.e.\ $\gamma_{\textrm{meas}} = \gamma_{\textrm{real}} L / c$. We note
that the parameters chosen accurately reflect those of real
microcavities at optical frequencies \cite{pcsel}; the complete set of
simulation parameters is given in Appendix D.

As shown in Fig.~\ref{fig1}, we find close agreement between SALT
calculations and FDTD simulations.  At a representative pump strength
$D_0' = 0.488 D_c$, the mode intensities produced by SALT differ from
those of the four- and six-level FDTD simulations by $\sim 1 \%$,
while the frequencies differ by $< 0.1 \%$. The difference in mode frequencies
between SALT and FDTD also exists at the first lasing threshold, for
which an analytical value can be calculated.  There, we find that the
FDTD simulation has a $0.2 \%$ error in the first mode frequency,
while SALT has a $0.08 \%$ error; this error arises from the spatial
discretization of the cavity employed in both approaches \cite{1d}.
It is worth emphasizing that SALT treats the non-linearity to infinite order;
in the earlier work on the Maxwell-Bloch model \cite{tandy} it was shown
that for this same cavity the common cubic approximation for the non-linearity
fails both quantitatively and qualitatively.

These results demonstrate that so long as the system satisfies SIA,
the mapping between systems with an arbitrary number of levels to an
effective two-level system is nearly exact, and SALT is able to very
accurately determine the steady state properties of the cavity.  If
two cavities, each with an arbitrary number of levels, have the same
effective parameters $D_0'$ and $\gamma_\parallel'$, and otherwise
have the same polarization relaxation rate and atomic transition
frequency, the cavities are equivalent from the electromagnetic point
of view, and will have identical lasing properties.

The six-level simulations shown in Fig.~\ref{fig1} occupy the
non-linear parameter regime of Eq.~(\ref{gp1})-(\ref{gp2}),
i.e.\ $\gamma_\parallel'$ is a linear function and $D_0'$ a non-linear
function of $\mathcal{P}$. However, the unscaled modal intensity
leaving the cavity is still, to leading order, linear in
$\mathcal{P}$. This can be seen by rearranging (\ref{TSG2}), inserting
the expressions for $\gamma_\parallel'$ and $D_0'$, and noting that at
the end of the cavity the inversion is roughly independent of the pump
strength. This result is discussed further in Appendix C.

\begin{figure}[!h]
\centering
\includegraphics[width=0.7\textwidth]{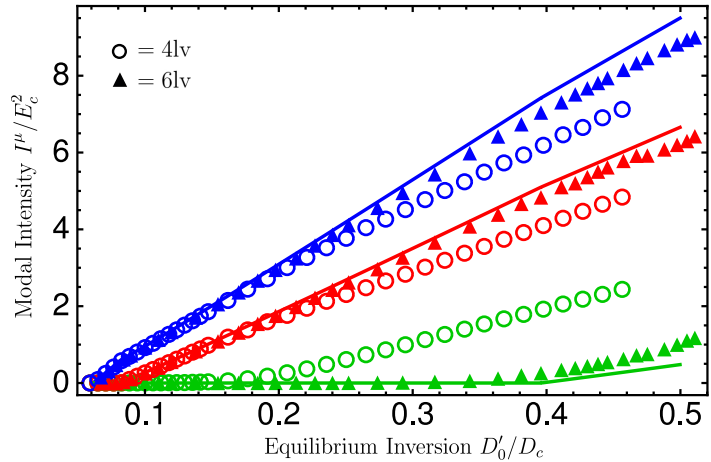}
\caption{Breakdown of the equivalence between SALT and FDTD when SIA
  is not valid is shown here in two different ways. Here,
  modal intensities as a function of the normalized equilibrium
  inversion $D_0' / D_{c}$ (effective pump) are shown for a 1D microcavity edge emitting laser
  with $\gamma_\perp = 4.0$ and $n=1.5$.  Solid lines again represent
  results obtained from SALT, while open circles represent FDTD
  simulations of a simple four-level system with $\gamma_\parallel' =
  0.1$. Triangles represent FDTD simulations of a
  six-level system in the non-linear parameter regime 
  in which $\gamma_\parallel' \sim 0.001$ for $D_0'
  \le 0.1$, and thus satisfying SIA, 
  but $\gamma_\parallel' \sim 0.01$ for $D_0' \ge 0.45$, and consequently
  no longer satisfying SIA. \label{fig2}}
\end{figure}

The mapping between the $N$-level laser and two-level SALT breaks down
at large pump strengths, when the condition $\gamma_\parallel' \ll
\gamma_\perp$ is violated due to the increase of $\gamma_\parallel'$
with $\mathcal{P}$.  In Ref.~\cite{tandy}, following an argument by
Haken \cite{fu}, it was demonstrated that violating this condition
causes the SIA for the two-level model to break down.  This effect can
be seen in the four-level laser data in Fig.~\ref{fig2}, where
$\gamma_\parallel' = 0.1,\gamma_\perp = 4.0$ and accuracy is already
lost for the third lasing mode.  For the six-level data of
Fig.~\ref{fig2}, which is in the non-linear parameter regime, 
SIA is satisfied and SALT agrees with the FDTD simulations for
small values of the normalized equilibrium inversion; for larger
values of $D_0'$, the SALT and FDTD results begin to diverge.

\begin{figure}[!h]
\centering
\includegraphics[width=0.7\textwidth]{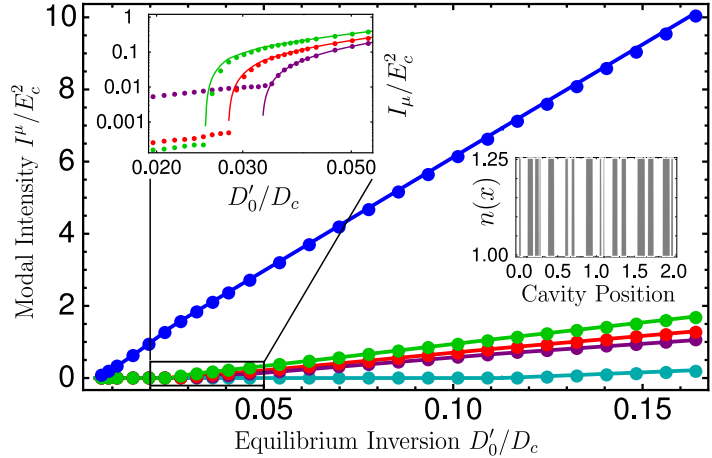}
\caption{SALT and FDTD results for a 1D random laser.  Modal
  intensities are plotted against the normalized equilibrium inversion
  $D_0' / D_{c}$ (effective pump).  Solid lines represent SALT results, and circles
  represent FDTD simulations for a four-level system with
  $\gamma_\parallel' = 0.001$.  The refractive index distribution of
  the edge emitting random laser is described in the text.  The gain
  medium has $\gamma_\perp = 4.0$ and is in the regime described by
  Eq.~(\ref{l1})-(\ref{l2}).  Left inset: log-log plot of the
  indicated region where three modes turn on in close proximity.
  Right inset: schematic of the cavity structure. \label{fig3}}
\end{figure}

Finally, to demonstrate that the mapping to an effective two-level
model works equally well for a complex laser cavity, Fig.~\ref{fig3}
shows a comparison between SALT and FDTD simulations for a four-level
gain medium in a 1D random dielectric structure. A number of studies
have been published on random lasers using such simulations
\cite{1drl1,1drl2}; SALT provides a much more efficient method for
such studies, which often require generating a statistical ensemble of
lasers.  Here, the passive cavity dielectric function contains $\sim
31$ layers, alternating randomly between regions with refractive
indices $n_1 = 1.25$ and $n_2 = 1$. Each random layer was generated
according to the formula $d_{1,2} = \langle d_{1,2} \rangle (1 + \eta
\zeta)$ where $\langle d_1 \rangle = (1/3)(L/30)$ and $\langle d_2
\rangle = (2/3)(L/30) $ are the average thicknesses of the layers,
$\eta = 0.9$ represents the degree of randomness of the cavity, and
$\zeta \in [ -1, 1 ]$ is a randomly generated number.  The gain medium
was added uniformly to the entire cavity, and the coherent pump was
likewise uniform. The transition frequency was chosen such that $n_1
k_a L = 120$, corresponding to roughly $20$ wavelengths inside of the
cavity.  We find only small discrepancies between the SALT and FDTD
results, with $\sim 1.1 \%$ difference in the modal intensities.
These differences did not vary significantly between different
realizations of the random laser.

\section{Computational efficiency of SALT for N-level systems}
\label{efficiency section}

\begin{figure}[!ht]
\centering
\includegraphics[width=1\textwidth]{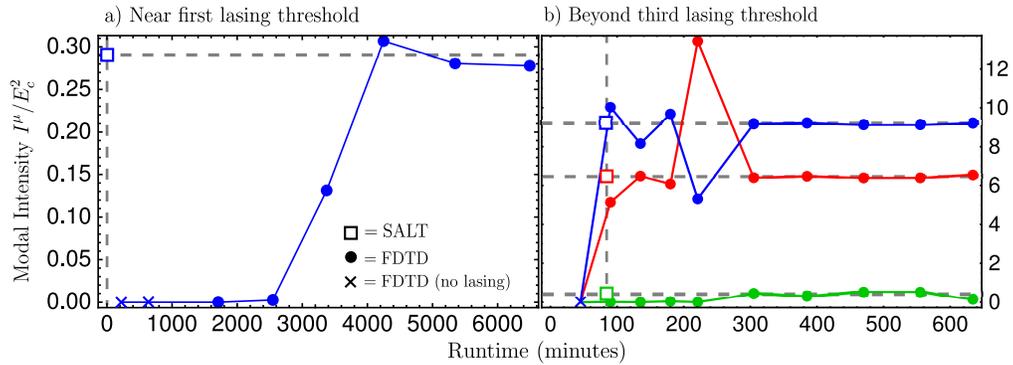}
\caption{Comparison of SALT and FDTD run-times.  Modal intensities are
  shown as a function of the run-time for SALT (squares) and
  four-level FDTD simulations (circles), using the parameters of
  Fig.~\ref{fig1}.  FDTD simulations that have not begun to lase are
  marked as crosses. Plot (a) shows data for $D_0/D_c = 0.071$, just
  above the first lasing threshold.  SALT determined the steady-state
  single modal intensity in under three minutes, while the FDTD
  required $\sim 5000$ minutes to reach steady state.  Plot (b) shows
  data for $D_0/D_c = 0.486$, well above the third lasing threshold.
  SALT calculated all data up to and including this pump value in
  under 90 minutes, whereas FDTD required $> 500$ minutes for the
  first two modes to reach steady-state, with the third mode intensity
  (green circles) still fluctuating after $5000$
  minutes (not shown).  \label{rtfig}}
\end{figure}

In this section we present a set of benchmarks comparing the
computational efficiency of SALT to FDTD.  SALT calculations enjoy
three main advantages over FDTD simulations of the semiclassical laser
equations.  First and foremost, SALT directly finds the steady-state
solutions, so no time integration is involved, which substantially
decreases computational effort.  Second, SALT unambiguously determines
how many modes are lasing at a given pump, whereas it can be difficult
to determine, especially for multimode lasing, when an FDTD simulation
has reached the steady-state with all modes that will lase ``on".
Third, within SALT, with minimal additional computational effort, it
is possible to monitor modes which are {\it below} threshold via a
modified threshold matrix \cite{saltsci}, and hence to ascertain if
more modes are likely to turn on in some interval of pump. It is important
to note that the implementation of SALT used in this study has
also not yet been fully optimized. For instance, the implementation
of SALT used in this study 
requires calculating the entire lasing intensity and frequency
spectrum starting from the first lasing threshold. However, this is
not necessary and it is possible to implement SALT to take an initial
guess of the number of modes and their relative intensities and then
allow the algorithm to flow to the correct solution as the SALT
algorithm has been demonstrated to be rather robust \cite{li_thesis}.
Thus, while one might assume from Fig.~\ref{rtfig} that there would
be a crossover pump value, high above threshold, where it would be
more efficient to calculate the steady-state solutions using FDTD,
this is not necessarily the case.

SALT does have one disadvantage that FDTD does not, assuming the cavity
is not in the chaotic regime. The convergence time in FDTD is
determined by the longest time scale in the problem, which is the greater of
beating period between consecutive modes and the relaxation oscillation time. 
These time scales are relatively
independent of the number of modes lasing in the cavity, so the
efficiency is largely independent of the pump in the multimode regime.
This is not the case for SALT, as the computational time
increases as $N^2$ where $N$ is the number of lasing modes. However,
as we see in Fig.~\ref{rtfig}, even a non-optimal implementation
of SALT is substantially more efficient than FDTD even when
calculating the steady state of a single pump value.

Calculating full modal intensity/frequency curves as a function of the
pump strength, such as in Fig.~\ref{fig1}, is generally much more
efficient using SALT.  For example, in order to generate the curves
seen in Fig.~\ref{fig1}, SALT ran for a little under 2 processor
hours.  To generate all of the FDTD data for the four level
simulations took 267 processor days.  If one is attempting to explore
a large parameter space of designs or system parameters, SALT may make
studies feasible which are simply impractical using FDTD, particularly
in more realistic 2D and 3D structures.

As mentioned before, the bulk of the computational effort required
for the SALT algorithm, especially in higher dimensions, is in solving
for the TCF states. While the difficulty of solving for the TCF states
does scale with the dimensionality of the system, one only need
solve the associated generalized eigenvalue problem $\sim 100$ 
(even in higher dimensions) to have a sufficiently complete basis
for all pump values, whereas in FDTD one needs to solve an $O(n^d)$
problem at each of many thousands, if not millions, of time steps.
Furthermore, it is likely that switching to a finite element method
in higher dimensions is not only possible, but likely to result in
increased computational efficiency. Once one has a TCF basis library,
using the SALT algorithm to iterate above threshold does not
directly scale with the dimensionality of the system. Finally, while
current implementations of SALT assume the electric field is
perpendicular to the direction of wave propagation, the vectorial
generalization of SALT is under investigation.

\section{Conclusion}

We have found that using stationarity conditions on the non-lasing
level populations, the rate equations for an $N$-level laser can be
mapped to an effective two-level model, for which the steady-state
multimode lasing properties are efficiently solvable using
Steady-state Ab initio Laser Theory (SALT).  Using this mapping, we
found excellent agreement between SALT and FDTD simulations for the
modal frequencies, thresholds and above-threshold intensities, in the
expected domain of validity.  SALT is typically several orders of
magnitude more efficient computationally than time domain solution of
the laser-rate equations, assuming only steady-state properties are
needed.

\section*{Acknowledgments}

We thank Robert J. Tandy, Hui Cao, and Peter Bermel for helpful
discussions.  This work was partially supported by NSF grant
No. DMR-0908437.

\appendix
\section{Classical polarization in the laser-rate equations}

Throughout this paper, we have used the density matrix equations of
motion for the polarization field.  However, much of the literature
uses the classical oscillating dipole equation, in which the gain
atoms are assumed to be dipoles undergoing harmonic oscillations
and the quantum density matrix is neglected.  
This appendix briefly
demonstrates that the two models are equivalent, and derives the
relevant parameter redefinitions.  A more thorough discussion has been
given by Boyd \cite{boyd}.

The polarization equation used here,
\begin{equation}
  \dot{P}^{+} = -\left(i\omega_a + \gamma_{\perp}\right)P^{+} +
  \frac{g^2}{i \hbar}E^+\left(\rho_{uu} - \rho_{ll}\right)
\end{equation}
is derived from the density matrix equation
\begin{equation}
  \dot \rho_{ul} = \left( i \omega_a + \gamma_\perp \right) \rho_{ul} - \frac{i}{\hbar} g E
  (\rho_{uu} - \rho_{ll}) \label{dme}
\end{equation}
where $|u \rangle$ and $|l \rangle$ are the upper and lower lasing
states, $\rho_{ij}$ is the density matrix element $ij$ and $g_{ul} =
g_{lu} = g$ is the coupling constant of the lasing states to the
electric field, which allows for the definition $P^+ = g
\rho_{ul}$. Alternatively, following the notation of Boyd \cite{boyd},
we could consider the equation of motion for $M = g\rho_{ul} +
\textrm{c.c.} = P^+ + P^-$, which is the expectation value of the
dipole moment induced by the applied field, i.e.\ the classical
oscillating dipole field. Thus
\begin{equation}
\dot M = g \dot \rho_{ul} + \textrm{c.c.} =
\left( i \omega_a + \gamma_\perp \right) \rho_{ul} - \frac{i}{\hbar} g E
D + \textrm{c.c.},
\end{equation}
using (\ref{dme}), and
\begin{equation}
  \ddot M = \left(-\omega_a^2 + 2i\omega_a \gamma_\perp +
  \gamma_\perp^2 \right) g \rho_{ul} - \frac{\omega_a}{\hbar}g^2 E D +
  \textrm{c.c.}.
\end{equation}
This can be rewritten as
\begin{equation}
\ddot M + 2 \gamma_\perp \dot M + \omega_a^2 M = -\gamma_\perp^2 M - \frac{2\omega_a g^2}{\hbar} ED,
\end{equation}
where $D = \rho_{uu} - \rho_{ll}$ is the inversion density of the
lasing states.  For $\omega_a^2 \gg \gamma_\perp^2$, we can discard
the term $\gamma_\perp^2 M$, resulting in the traditional form of the
classical oscillating dipole polarization field equation.  This gives
the classical coupling constant $\sigma = 2 \omega_a g^2 / \hbar $
\cite{siegman}.

A similar analysis can used to show that the inversion equation can be rewritten as
\begin{equation}
\dot D = -\gamma_\parallel(D - D_0) + \frac{2 E}{\hbar \omega_a} \dot M,
\end{equation}
which demonstrates how the change in the inversion is dependent upon the classical
polarization field.  Therefore, so long as $\omega_a^2 \gg \gamma_\perp^2$, a condition that is usually satisfied,
these two formulations of the polarization dynamics are equivalent.

\section{Effective two-level parameters from N-level rate equations}

This appendix derives the effective equilibrium inversion and
relaxation rate for a gain medium with an arbitrary number of levels.
We allow decays between any two levels, even if they are not adjacent.
We assume there is a single lasing transition, between levels
$|u\rangle$ and $|l\rangle$, which need not be adjacent.  The rate
equation for an arbitrary \textit{non-lasing} level in the system is
\begin{equation}
  \dot \rho_{ii} = \sum_{j} \gamma_{ij}\rho_{jj} - \sum_{j}
  \gamma_{ji} \rho_{ii} + \gamma_{iu} \rho_{uu} + \gamma_{il}
  \rho_{ll} - (\gamma_{ui}+\gamma_{li})\rho_{ii},
  \label{non lasing rhodot}
\end{equation}
where the summations are taken over all non-lasing levels.  Here we do
not distinguish between decay rates and pumping rates; $\gamma_{ij}$
is simply interpreted as the rate at which level $|j \rangle$
transitions into level $|i \rangle$, regardless of the energies of
those states.  If the populations of all the non-lasing transitions
are stationary, i.e.~$\dot \rho_{ii} = 0$, then we can rewrite
(\ref{non lasing rhodot}) as
\begin{eqnarray}
  \sum_j R_{ij} \rho_{jj} &=& \gamma_{iu} \rho_{uu} +
  \gamma_{il}\rho_{ll}, \label{genrate} \\ R_{ij} &\equiv& (s_i +
  \gamma_{ui} + \gamma_{li}) \delta_{ij} - \gamma_{ij}.
\end{eqnarray}
Here, $s_i \equiv \Sigma_j \gamma_{ji}$ and $\delta_{ij}$ is the
Kronecker delta.  Inverting (\ref{genrate}) gives
\begin{equation}
  \rho_{ii} = \sum_j [R^{-1}]_{ij} (\gamma_{ju} \rho_{uu} +
  \gamma_{jl} \rho_{ll}). \label{rhoii}
\end{equation}
Hence, we can express the total number density of gain atoms as
\begin{eqnarray}
  n &=& \sum_i \rho_{ii} + \rho_{uu} + \rho_{ll} \\ &=& T_u \rho_{uu}
  + T_l \rho_{ll}
\end{eqnarray}
where
\begin{eqnarray}
  T_u &=& 1 + \sum_{ij} [R^{-1}]_{ij} \gamma_{ju}, \\
  T_l &=& 1 + \sum_{ij} [R^{-1}]_{ij} \gamma_{jl}.
\end{eqnarray}
Noting that $D = \rho_{uu} - \rho_{ll}$, we can write the populations
of the lasing states as
\begin{eqnarray}
  \rho_{uu} &=& \frac{n + T_l D}{T_l + T_u}, \label{rhouu}\\
  \rho_{ll} &=& \frac{n - T_u D}{T_l + T_u}. \label{rholl}
\end{eqnarray}

From the equations of motion for the lasing levels, we have the
inversion equation
\begin{equation}
  \dot D = \dot \rho_{uu} - \dot \rho_{ll} = \sum_i(\gamma_{ui} -
  \gamma_{li})\rho_{ii} - s_u \rho_{uu} + s_l \rho_{ll} -
  \frac{2}{i\hbar}\mathbf{E}^+ \cdot \left((\mathbf{P}^{+})^* -
  \mathbf{P}^{+} \right),
\end{equation}
where $s_u = \Sigma_j \gamma_{ju} + \gamma_{lu}$ and $s_l$ is defined
similarly.  Inserting (\ref{rhoii}) into this equation gives
\begin{equation}
  \dot D = B_u \rho_{u,u} + B_l \rho_{l,l} -
  \frac{2}{i\hbar}\mathbf{E}^+ \cdot \left((\mathbf{P}^{+})^* -
  \mathbf{P}^{+} \right),
\end{equation}
where
\begin{eqnarray}
  B_u &\equiv& - s_u + \sum_{ij} (\gamma_{ui} - \gamma_{li}) [R^{-1}]_{ij}
  \gamma_{ju}, \\ B_l &\equiv& s_l + \sum_{ij} (\gamma_{ui} -
  \gamma_{li}) [R^{-1}]_{ij} \gamma_{jl}.
\end{eqnarray}
Plugging in (\ref{rhouu}) and (\ref{rholl}) now yields the inversion
equation in the desired form (\ref{efinv}), with
\begin{eqnarray}
\gamma_\parallel' &=& \frac{B_l T_u - B_u T_l}{T_u + T_l}, \\
D_0' &=& \frac{B_u + B_l}{B_lT_u - B_uT_l} n.
\end{eqnarray}

\section{Inversion as a function of the pump}

\begin{figure}[!ht]
\centering
\includegraphics[width=1\textwidth]{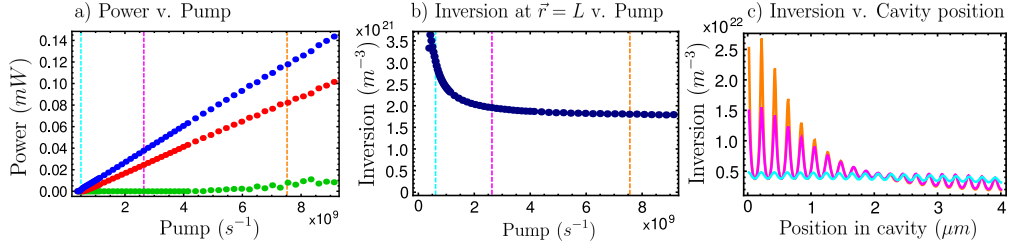}
\caption{(a) Unscaled modal intensity of the six-level simulations
  from Fig.~\ref{fig1} as a function of the pump. A cross sectional
  area of $1 \textrm{m}^2$ is assumed to calculate the power.
  (b) Inversion as a
  function of the pump at the cavity boundary. Dashed lines in plots
  a and b correspond to the pump values shown in plot c.
  (c) Inversion as a
  function of position within the cavity for three different pump
  values, cyan corresponds with $\mathcal{P} = 3.75 \times 10^8 s^{-1}$, magenta with
  $\mathcal{P} = 1.65 \times 10^9 s^{-1}$, and orange with $\mathcal{P} = 4.85 \times 10^9 s^{-1}$ to show the
  evolution of the inversion within the cavity as a function of the pump strength.    \label{intvpump}}
\end{figure}

In this appendix we discuss the surprising result that the unscaled modal
intensities of the six-level simulations discussed in Fig.~\ref{fig1} 
as a function of the pump are approximately linear, as
shown in Fig.~\ref{intvpump}, even though these six-level simulations
are in the non-linear parameter regime. In simple treatments of lasers
\cite{siegman} the inversion is assumed to be clamped after the first
lasing threshold. However, a more detailed analysis lead to the inclusion
of spatial hole-burning effects which forces the inversion
in the cavity
to change beyond the first lasing threshold in a non-linear manner.
This result then, that the unscaled modal intensity is linear in the
pump rate, can be understood from the observation that at the
end of the cavity, $\vec r = L$, all of the lasing modes have a maximum
in their fields, and thus the inversion is effectively clamped at this
point beyond the first lasing threshold, which can be seen in Fig.~\ref{intvpump}
(b).

To understand this formally, we begin by rewriting (\ref{TSG2}) and removing the
scaling factors $E_c$ and $D_c$ gives
\begin{equation}
\frac{2g^2}{\hbar^2} \sum_{\nu=1}^N \Gamma_\nu |\Psi_\nu(\vec r)|^2 = \gamma_\parallel' \left(
\frac{D_0'}{D(\vec r)} -1 \right).
\end{equation}
Substituting in (\ref{gp1}) and (\ref{gp2}), which are valid for this
simulation, gives
\begin{equation}
\frac{g^2}{\hbar^2}\sum_{\nu=1}^N \Gamma_\nu |\Psi_\nu(\vec r)|^2 = \mathcal{P} \left(\frac{N}{D(\vec r)} -1
\right) - \gamma_{12}.
\end{equation}
The inversion $D$ is a function of both position and the pump.
However, for $\vec{r}=L$ corresponding to the cavity edge, $D$ should be
mostly independent of the pump, as at this location every mode is at its
maximum intensity and the effect of spatial hole-burning is most
pronounced.  The FDTD simulation results, shown in
Fig.~\ref{intvpump}, demonstrate that $D$ indeed varies very weakly
with $\mathcal{P}$ at the cavity edge.

\section{Simulation constants}

In this appendix we list the parameters used in each of the FDTD
simulations described above.  These constants are matrices, with
$\gamma_{ij}$ denoting the decay rate from $|j \rangle$ to $|i
\rangle$.  These values are given in their dimensionless form,
i.e.\ $\gamma_{\textrm{meas}} = \gamma_{\textrm{real}} L / c$.
Unlisted entries are zero. We also note that throughout this paper $|0
\rangle$ denotes the ground state, so these matrices are $0$ indexed.

For the four-level simulations in Fig.~\ref{fig1},
\begin{equation}
\gamma_{4\textrm{lv, fig \ref{fig1}}} = \left( \begin{array}{cccc}
  \cdot& 0.8 & \cdot& \cdot\\
  \cdot& \cdot& 5 \times 10^{-4} & \cdot\\
  \cdot& \cdot & \cdot& 0.8 \\
  \cdot& \cdot& \cdot& \cdot \\
\end{array} \right).
\end{equation}
The dipole matrix element is $g = 2.3 \cdot 10^{-12}
\textrm{m}^{3/2}$, and the number of gain atoms is $n = 5
\cdot 10^{23} \textrm{m}^{-3}$.  The pump $\mathcal{P}$ was varied
between $3 \cdot 10^{-6}$ and $3 \cdot 10^{-5}$.

Thus, for an optical wavelength of $\lambda = 628\textrm{nm}$, the
requirement in Fig.~\ref{fig1} that $n_0kL = 60$ means that $L = 4 \mu
\textrm{m}$. Using this length, the decay rates can be converted to
their unit-full values as $\gamma_\perp = 3 \cdot 10^{14}
\textrm{s}^{-1}$, $\gamma_{23} = \gamma_{01} = 6 \cdot 10^{13}
\textrm{s}^{-1}$, $\gamma_{12} = 3.75 \cdot 10^{10} \textrm{s}^{-1}$,
and the pump at threshold is $\mathcal{P} = 3 \cdot 10^8
\textrm{s}^{-1}$. Similarly, the dipole matrix element also acquires
units of inverse time, and can be expressed as $g^2 / \hbar = 3.98
\cdot 10^{-9} \textrm{m}^3 / \textrm{s}$, which corresponds to a
coupling constant in the classical oscillating dipole picture of
$\sigma = 10^{-4} \textrm{C}^2/\textrm{kg}$. These constants can be
seen to be similar to those used in other studies of optical
microcavities \cite{york,pcsel}.

For the six-level simulations in Fig.~\ref{fig1},
\begin{equation}
\gamma_{6\textrm{lv, fig \ref{fig1}}} = \left( \begin{array}{cccccc}
\cdot& 0.8 & 10^{-5} & 10^{-5} & 10^{-5} & 10^{-5} \\
\cdot& \cdot& 0.8 & 10^{-5} & 10^{-5} & 10^{-4} \\
\cdot& \cdot& \cdot&  5 \times 10^{-5} & 10^{-5} & 10^{-5} \\
\cdot& \cdot& \cdot& \cdot& 0.8 & 10^{-5}\\
\cdot& \cdot& \cdot& \cdot& \cdot& 0.8\\
\cdot& \cdot& \cdot& \cdot& \cdot& \cdot\\
\end{array} \right).
\end{equation}
Furthermore, $\gamma_{15} = 10^{-4}$, and the lasing transition is
between levels $|3 \rangle$ and $|1 \rangle$ (where the ground state
is again $|0 \rangle$ and the states are numbered in order of
increasing energy).

For the four-level simulations in Fig.~\ref{fig2},
\begin{equation}
\gamma_{4\textrm{lv, fig \ref{fig2}}} = \left( \begin{array}{cccc}
\cdot& 0.8 & \cdot&\cdot \\
\cdot& \cdot& 5 \times 10^{-2} & \cdot\\
\cdot& \cdot& \cdot& 0.8 \\
\cdot& \cdot& \cdot&\cdot \\
\end{array} \right).
\end{equation}

For the six-level simulations in Fig.~\ref{fig2},
\begin{equation}
\gamma_{6\textrm{lv, fig \ref{fig2}}} = \left( \begin{array}{cccccc}
\cdot& 0.8 & 10^{-5} & 10^{-5} & 10^{-5} & 10^{-5} \\
\cdot&\cdot & 0.8 & 10^{-5} & 10^{-5} & 10^{-4} \\
\cdot&\cdot & \cdot&  5 \times 10^{-5} & 10^{-5} & 10^{-5} \\
\cdot&\cdot & \cdot&\cdot & 0.8 & 10^{-5}\\
\cdot&\cdot & \cdot&\cdot &\cdot & 0.8\\
\cdot&\cdot & \cdot&\cdot &\cdot &\cdot \\
\end{array} \right).
\end{equation}

The four-level simulations of the random cavity in Fig.~\ref{fig3}
used the same parameters as the four-level simulations in
Fig.~\ref{fig1}.


\begin{thebibliography}{99}

\bibitem{haken} H. Haken, \textit{Light: Laser Dynamics} Vol. 2 (North-Holland
Phys. Publishing, New York, 1985).

\bibitem{lamb63} { W. E. Lamb, ``Theory of an Optical Maser,'' Phys. Rev. \textbf{134},
A1429 (1964).}

\bibitem{siegman} A. E. Siegman, \textit{Lasers} (University Science Books, Mill Valley - 
California, 1986).

\bibitem{york} A. S. Nagra and R. A. York, ``FDTD analysis of wave propagation in nonlinear
absorbing and gain media,'' IEEE Trans. Antennas Propag. \textbf{46}, 334-340 (1998).

\bibitem{yee} K. S. Yee, ``Numerical solution of the initial boundary value problems
involving Maxwell's equations in isotropic media,'' IEEE Trans Antennas Propag. \textbf{14},
302-307 (1966).

\bibitem{salt1} H. E. T\"{u}reci, A. D. Stone, and B. Collier, ``Self-consistent
multimode lasing theory for complex or random lasing media,'' Phys. Rev. A
\textbf{74}, 043822 (2006).

\bibitem{salt2} H. E. T\"{u}reci, A. D. Stone, and L. Ge, ``Theory of the spatial structure
of nonlinear lasing modes,'' Phys. Rev. A \textbf{76}, 013813 (2007).

\bibitem{saltsci} H. E. T\"{u}reci, L. Ge, S. Rotter, and A. D. Stone, ``Strong interactions
in multimode random lasers,'' Science \textbf{320}, 643-646 (2008).

\bibitem{spasalt} L. Ge, Y. D. Chong, and A. D. Stone, ``Steady-state ab initio laser
theory: generalizations and analytic results,'' Phys. Rev. A \textbf{82}, 063824 (2010).

\bibitem{cao} H. Cao, ``Review on the latest developments in random lasers with coherent
feedback,'' J. Phys. A \textbf{38}, 10497-10535 (2005).

\bibitem{phot} O. Painter, R. K. Lee, A. Scherer, A. Yariv, J. D. O'Brien, P. D. Dapkus,
and I. Kim, ``Two-dimensional photonic band-gap defect mode laser,'' Science \textbf{284},
1819-1821 (1999).

\bibitem{pcsel} S. Chua, Y. D. Chong, A. D. Stone, M Solja\u{c}i\'{c}, and J. Bravo-Abad,
``Low-threshold lasing action in photonic crystal slabs enabled by Fano resonances,'' Opt.
Express \textbf{19}, 1539 (2011).

\bibitem{disk} C. Gmachl, F. Capasso, E. E. Narimanov, J. U. N\"{o}ckel, A. D. Stone, 
J. Faist, D. L. Sivco, and A. Y. Cho, ``High-power directional emission from microlasers
with chaotic resonators,'' Science \textbf{280}, 1556-1564 (1998).

\bibitem{li_thesis}  Li Ge, Yale PhD thesis, 2010.

\bibitem{tandy} L. Ge, R. J. Tandy, A. D. Stone, and H. E. T\"{u}reci, ``Quantitative verification
of ab initio self-consistent laser theory,'' Opt. Express \textbf{16}, 16895 (2008).

\bibitem{TE} The equations are written for the TM case, the modifications for TE are straightforward.

\bibitem{caveat} The observation that coherent and incoherent pumping are nearly equivalent for most systems, is
invalid when the coherent pumping is supplied at a similar frequency
to the atomic lasing transition and thus interactions between the lasing
field and pumping field must be taken into account.  

\bibitem{fu} H. Fu and H. Haken, ``Multifrequency operations in a short-cavity standing-wave
laser,'' Phys. Rev. A \textbf{43}, 2446-2454 (1991).

\bibitem{khanin} Y. I. Khanin, \textit{Principles of Laser Dynamics} (Elsevier, Amsterdam,
1995).

\bibitem{bid} B. Bid\'{e}garay, ``Time discretizations for Maxwell-Bloch equations,''
Numer. Meth. Partial Differential Equations \textbf{19}, 284-300 (2003).

\bibitem{1d} For any 1D cavity which is uniformly pumped the TCF states for solving SALT can also be found using a
transfer matrix method which does not require discretizing space.  We use a more general TCF solver in the calculations
presented here which does discretize space.

\bibitem{1drl1} X. Jiang and C. M. Soukoulis, ``Time Dependent Theory for Random Lasers,'' Phys. Rev. Lett.
\textbf{85}, 70 (2000).

\bibitem{1drl2} X. Jiang, S. Feng, C. M. Soukoulis, J. Zi, J. D. Joannopoulos, and H. Cao, ``Coupling,
competition, and stability of modes in random lasers,'' Phys. Rev. B \textbf{69}, 104202 (2004).

\bibitem{boyd} R. W. Boyd, \textit{Nonlinear Optics} (Academic Press, New York, 2008).

\end{thebibliography}
\end{document}